# FRAILTY INDEX AS A MAJOR INDICATOR OF AGING PROCESSES AND MORTALITY IN ELDERLY:

## Results From Analyses of the National Long Term Care Survey Data*

A. Kulminski, A. Yashin, I. Akushevich, S. Ukraintseva, K. Land, K. Arbeev, and K. Manton.

(Center for Demographic Studies, Duke University)

**ABSTRACT**

To better understand mortality change with age capturing the variability in individuals' rates of aging, we performed comprehensive analysis of statistical properties of a cumulative index of age-associated disorders (deficits), called a "frailty index" (FI). This index is calculated as the proportion of the health deficits in an individual. It is found, first, that frequency, time-to-death, mortality-rate, and relative-risk-of-death exhibit remarkably similar FI- and age- patterns. Second, the FI, on the one hand, and mortality rate and relative risk, on the other hand, also exhibit similar age patterns with accelerated increase up to oldest-old ages and with subsequent deceleration and even decline. Third, distribution of the FI with time-to-death is sharper than that of age with time-to-death. These and related findings support the conclusion that the FI can describe aging processes and population heterogeneity. We also discuss the ability of the FI to capture physiological processes underlying aging both on individual and population levels.

*The research reported here was supported in part by P01-AG-017937-05 grant from the National Institute on Aging (NIA). A.K. also acknowledges support from K12-AG-000982-05 NIA grant.

# INTRODUCTION

The increase of human life expectancy in developed countries has lead to considerable growth of elderly populations. In 2003 in the U.S., there were 35.9 million elderly (65+) individuals – 12.3% of the total U.S. population. It is expected that, by 2030, this number will increase to 71.5 million – roughly 20% of the U.S. population. While the populations in the world become older, neither life expectancy nor chronological age (i.e., time passed from the date of birth) directly measure aging processes. Indeed, individuals do not age with the same rate. Certain individuals remain alive to extreme ages being in good health and functioning well. Other individuals survive to advanced ages despite health or function decline (Evert et al. 2003). Others have poor health/quality of life characteristics and die early. Therefore, neither life expectancy nor chronological age captures physiological basis underlying aging processes. Consequently, they are not good indicators of aging either on a population or individual level.

Kowald (2002) argued that instead of life expectancy, mortality risk is a better indicator of aging. Mortality risk is a population characteristic which reflects and summarizes individual courses of life, i.e., individual aging processes (Jazwinski 2002). However, this appears to be more suitable characteristic for short-lived experimental species for which birth and death can be easily followed. For humans, it is better to have more quickly responding characteristic not to follow individuals from their birth do death. These rationales argue search for better indicator of aging processes (at least in humans) which could reflect the underlying physiological basis. Actually, biological or physiological processes on an individual level constitute the basis for the concept of biological age (BA), which may better predict mortality and survival (Bennett 2004).

On the other hand, researches intended to capture the variability in an individuals' rate of aging adjusting models describing mortality on a population level introducing formal concepts. Beard (1971) introduced the concept of longevity summarized in the longevity factor. Later, Vaupel et al. (1979) introduced the concept of individual frailty. However, these concepts were primarily needed for the development of the demographic models of ageing and mortality and were not intended to be based on specific physiological processes in an aging organism.

Studies of the individual's BA have significant potential to fill these formal constructions with physiological meaning. One such conception which underlies BA and captures population heterogeneity is related to the phenomenon of physiological frailty (Fried et al. 2004). Frailty refers to a specific physiological state that is not necessarily associated with chronic conditions or disability and that typically arises in the elderly. Despite numerous efforts to define the concept of frailty in the elderly, there is still a lack of a standard operational definition which is well suited for epidemiologists and clinicians (Rockwood, Hogan, and MacKnight 2000; Rockwood, Song, and MacKnight 2005; Bortz 2002; Gillick 2001; Fried et al. 2001; Morley, Perry, and Miller 2002). Nevertheless, because of population aging, the American Medical Association expresses concerns about the growing population of frail older individuals (Council 1990) stating that almost 40% of individuals aged 80+ are "frail".

According to the current view, frailty can be considered as a physiological state of increased vulnerability to stressors that results from decreased physiological reserves, and potentialy disregulation of multiple physiologic systems. This decreased reserve results in difficulty maintaining homeostasis in response to "normal" perturbations that would not create such problems at younger ages (Fretwell 1993; Buchner 1992; Lipsitz and Goldberger 1992; Fried and Walston 2003; Yates 2002). Frailty appears to be multidimensional affecting various systems of an aging organism (Puts, Lips, and Deeg 2005). Nevertheless, heavy preference in prior studies was given to the use of unidimensional approaches to capture biomedical processes (Markle-Reid and Browne 2003; Hogan, MacKnight, and Bergman et al. 2003).

Since frailty appears to be a consequence of decline in physiological functions on an individual level, it might be captured by measuring various signs, symptoms, and abnormalities associated with health and/or quality of life, which could reflect these changes in an individual. This is the underlying paradigm of the approach proposed by Rockwood and Mitnitski with colleagues (Mitnitski et al. 2002; Mitnitski, Song, and Rockwood 2004; Rockwood et al. 2000; Rockwood, Mitnitski, and MacKnight 2002; Rockwood, Mogilner, and Mitnitski 2004) to describe frailty using conventional data available not only in clinical but also in population-based studies and surveys. These signs, symptoms, and abnormalities are commonly referred to as "deficits" and constitute the basis for construction of a frailty index (FI) calculated as the proportion of the health deficits in an individual. The FI may have the ability to characterize individual physiological frailty using survey data. The FI as a mean accumulation of deficits can predict death and describe health risks. The FI appears to be robust construct which is independent of nature of deficits. Consequently, it might be a candidate for the efficient expression of BA (Rockwood et al. 2002), which is recognized as a particular need to better understand the contribution of genetic heterogeneity to variations in human longevity (Kirkwood 2002). Such approach, determining a FI using self-reports on various deficits, appears promising to assess physiological frailty, capture hidden population heterogeneity, provide physiological basis for heterogeneity variables in models of mortality and aging, and capture individual and population BA.

In this study we adapted the Rockwood-Mitnitski's approach to constructing FI using National Long Term Care Survey (NLTCS) data. The NLTCS is a nationally-representative, longitudinal survey assessing health and functioning of the U.S. elderly (65+) individuals over nearly two decades (currently 1982 to 1999). To define the FI we will use the same, or similar, health deficits which were assessed using the Canadian Study of Health and Aging (CSHA) (Mitnitski, Mogilner, and Rockwood 2001). We comprehensively address three aspects which were not adequately studied in early works on FI. Specifically, our primary goal is to provide solid arguments that FI can describe population heterogeneity, predict death in elderly populations, and, consequently, be an indicator of BA suitable for understanding aging processes and mortality in population based studies. Second, we focus our analysis on disabled individuals who is at excessive risk of death, i.e., for whom all these aspects can be better pronounced. Third, we comprehensively address the problem of sex-sensitivity of the FI in the disabled elderly. The importance of the problem of disabled elderly is stressed by the fact that these individuals are more likely to be frail than non-disabled individuals. Since the impact of frailty on health/quality-of-life status characterized by the FI is likely to be of nonlinear nature, i.e., sensitive to the FI magnitude (Kulminski et al. 2005), focus on the disabled elderly would be even more important.

## METHODS
### Data
The NLTCS data were gathered during five surveys performed in 1982, 1984, 1989, 1994, and 1999. The NLTCS uses a sample of individuals drawn from national Medicare enrollment files. The survey instruments used in all five NLTCS waves asked the same disability, functional, and medical condition questions in the same way to minimize cross survey bias of estimates. The likelihood of this bias is also reduced by the high (95%) response rates in all NLTCS waves. The NLTCS samples contain longitudinal and cross sectional components. A two-stage interviewing process was used for selection of the NLTCS participants. First, a screening interview assessing chronic disability is given to all members of the sample (roughly, 20,000 for each wave) – except persons who received a community or institutional detailed interview in a prior NLTCS. Those persons reporting on the screen at least one impairment in an (Instrumental) Activity of Daily Living, (I)ADL, that had lasted, or was expected to last, 90+ days were then given either an in-person detailed community or institutional interview. Individuals, who, in a prior NLTCS, received such an interview, were automatically given a detailed questionnaire (of the appropriate type determined according to Census criteria on housing type at the time of the personal interview) in the current NLTCS (up to the time of death). To replace deceased individuals and to ensure that the sample screened is representative of the entire U.S. elderly population, a new sample supplement is drawn for each survey of persons who passed age 65 since the last NLTCS. All NLTCS records are linked to Medicare (to the end of 2001) and Vital Statistics (to August 6, 2003) files. The 1982 to 1999 NLTCS screener questionnaires represent roughly 42,000 different individuals. Detailed information in community surveys was gathered from about 26,700 interviews in all five NLTCS waves.

In the 1994 NLTCS wave an additional sub-sample of 1,762 "healthy" persons ("healthy" supplement [HS]) was selected. These persons were designated to receive a detailed interview even if screened initially as non-institutional unimpaired. The HS for the 1999 wave includes surviving persons selected for the 1994 HS (1,262), persons newly selected from the replacement (aged-in) component of the 1999 sample (283), and persons newly selected from the longitudinal component of the 1999 sample who were screened out in 1989 and not selected for the 1994 wave (64), comprising a total of 1,545 persons in the 1999 HS.

**The Frailty Index (FI)**
To construct the FI, we selected the subset of deficits (32 questions), which are the most similar to those assessed from CSHA (Mitnitski et al. 2001): difficulty with eating, dressing, walk around, getting in/out bed, getting bath, toileting, using telephone, going out, shopping, cooking, light house work, taking medicine, managing money, arthritis, Parkinson's disease, glaucoma, diabetes, stomach problem, history of heart attack, hypertension, history of stroke, flu, broken hip, broken bones, vision problem, self-rated health, trouble with bladder or bowels, dementia, hearing problem, visit of hearing therapist, dentist, and foot doctor. All these questions are presented in all five waves. Following Mitnitski et al. (2001), we define the FI as an unweighted count of the number of such deficits divided by the total number of all potential deficits considered for a given person. For instance, if individual have been administered 30 questions and responded positively (there is a deficit) to five and negatively (no deficit) to 24 of them, then the FI for the given person will be equal to 5/29. In this way we avoid the problem of missing answers counting only those questions which were explicitly answered in the survey. The importance of the problem of missing data is also reduced by high overall response rate in the survey (for most of the questions the missing rate is less than 1%).

**Time To Death And Time Under Observation**
NLTCS data are linked with Medicare Vital Statistics data. This allows us to follow-up the survival status of all NLTCS participants to August 6, 2003, i.e., up to 21 years. Using these data we construct a variable which describe individual time to death (TTD) and time under observation. In general, these characteristics were calculated as the number of days during which an individual stays alive since the date of interview in the survey. However, due to the large proportion of missing data on the date of interview, especially at early waves, for those with a missing interview date we have chosen the mean date of the period of interview for each wave. Since the interviews were conducted during a short period of time (three-four months), this method of filling the missing interview data does not introduce significant error in our pattern estimates. This was verified by varying the date within specified range (three-four months) and calculating the resulting patterns. The likelihood of such errors is also reduced by averaging individual times within certain predetermined intervals, e.g., one year.

**RESULTS**
**Frequency Distributions**
The focus of this paper is on disabled individuals. Therefore, we selected subgroups of individuals who were eligible for the detail community questionnaire because of disabled conditions, i.e., persons not in the HS. To increase the statistical power and smooth our estimates we pooled data for all waves into one sample (denoted as PAD). The resulting PAD sample represents 24,213 individuals (not necessarily different) who completed interviews (8,249 or 34% of them by males) from 1982 to 1999. In our analysis, we do not consider longitudinal components of the survey (a separate study will focus on longitudinal analyses), i.e., we assume that all individual measurements are independent (in fact, considering each NLTCS wave we checked that there was no any systematical bias due to such assumption).

We begin our analysis by studying the frequency patterns of deceased and surviving individuals. FI frequency patterns were calculated as a proportion of individuals at given FI level (we distinguished FI deciles) and dying/surviving in one, four, and eight years after the date of interview (Figure 1, upper panels). We also selected five-year age groups (e.g., 65-69, 70-74, etc.) to calculate age frequency patterns (Figure 1, lower panels). Figure 1 shows similarities in the shapes of the FI and age frequency distributions for surviving as well as for deceased individuals. Meanwhile, there are qualitative differences in the shapes of the FI and age patterns between surviving and deceased individuals. Specifically, the frequency patterns for the survivors are positively skewed being close in shape to a gamma distribution, while those for deceased individuals are more symmetric being close in shape to a normal distribution.

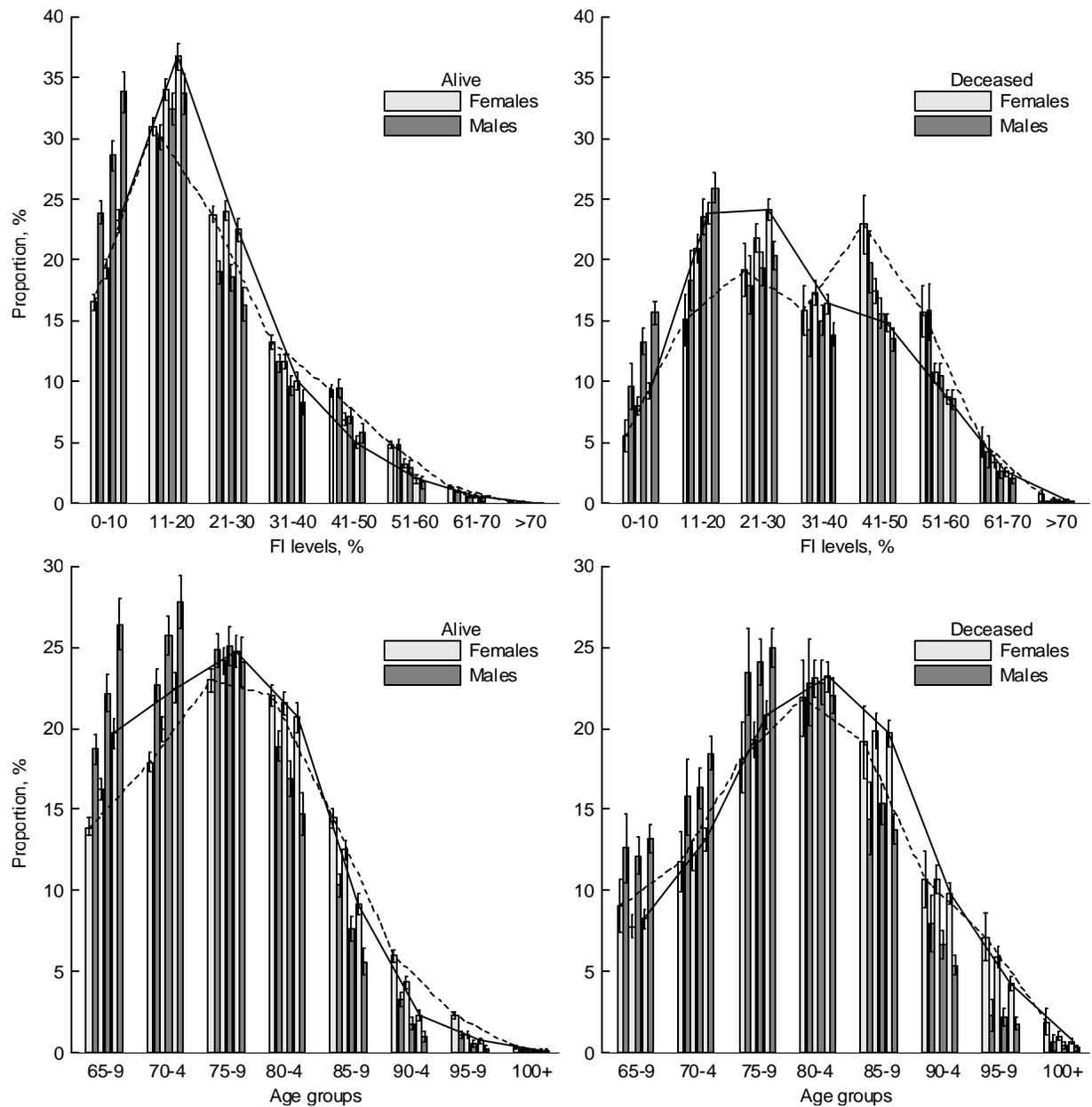

**Figure 1. Frailty Index (FI) (upper panels) and age (lower panels) frequency patterns for surviving (left panels) and deceased (right panels) males and females. Vital status is assessed during one, four, and eight years (left to right in**

**each group of bars) after the date of respective interview. Continuous (dashed) line shows representative shapes of the frequency patterns for eight (one) year vital status follow-up. Thin bars show 95% confidence interval (CI).**

Similarities between FI and age densities in PAD sample could be due to high correlation of the FI and age. Actually, although such correlation is statistically significant, this is clearly not the case since correlation coefficients between FI and age ($r_{FI-Age}$) are relatively small. Specifically, for the entire PAD sample we have $r_{FI-Age} = 0.193$, for males 0.127, and for females 0.221. Moreover, for deceased individuals these coefficients are even smaller. Therefore, we conclude that both age and FI have potential to characterize aging processes being rather independent measures. Peculiarities in the shapes of the FI and age frequency patterns, as well as the differences in their dynamics also support this conclusion.

Indeed, the shapes of the patterns for surviving males and females remain qualitatively similar irrespective of the observation time. Meanwhile, the FI patterns for deceased individuals experience change in shape from being negatively to positively skewed (while remaining close to a normal distribution) as observation time increases. This is not the case for age patterns which remain positively skewed with time. In all cases, as the observation time is increased, the number of survived individuals is redistributed toward smaller FI and younger ages. The shape of the FI frequency patterns and their change with time for decedents clearly show that FI is a useful characteristic for describing population heterogeneity in models of mortality and aging. Moreover, it provides necessary physiological background for frailty parameters in such models (Woodbury and Manton 1977; Yashin, Manton, and Vaupel 1985; Manton and Yashin, 2000; Manton, Akushevich, and Kulminski 2005), as discussed thereafter.

Both the FI and age patterns show that proportions of survivors at younger ages (69-74 years) and small FI (0-20%) are larger than decedents. Physiological basis for this is likely that survivors are healthier. Proportion of males for ages 69-74 years and FI of 0-10% is larger than females. This is likely a reflection of the fact that males have shorter life expectancy than females. Proportions of surviving males and females at younger ages and small FI increase as time under observation increases that supports the view that to survive during longer time individuals have to be healthier.

For larger FI levels and later ages (for 11-20% FI and 75-79 age group for survivors, and for 11-30% FI and 80-84 age group for decedents) proportion of females is similar to males and increases afterwards. Therefore, males need to be healthier than females to survive by advanced ages. The fact of convergence of the sex-specific age patterns also indicates that males and females age with different rates. An intriguing fact is that behavior of the FI patterns is similar to that of age patterns, i.e., FI patterns for males and females also converge. This similarity provides support toward the view that FI has a potential to characterize BA.

For large FI levels there is almost no difference in the proportions of male and female survivors, although prevalence of females at advanced ages is clearly pronounced irrespective of the vital status. Sex-insensitivity of the FI at its large values might indicate saturation effect – as health problems begin to affect majority of an organism's critical subsystems, sex becomes unimportant factor. Since the proportion of decedents at large FI levels is larger than the proportion of survivors, saturation by deficits will likely result in faster death. Nevertheless, the presence of survivors with large FI levels provides optimistic perspectives showing that individuals with large a FI can live quite long. One reason for that might be high stress-resistance of an organism in certain groups of individuals (Semenchenko et al. 2004).

Another intriguing observation is that there are fixed points in time for age and FI patterns for survivors (21-30% FI and 75-79 age group) and decedents (31-40% FI and 80-84 age group). At these points, the proportions of males and females remain nearly the same irrespective of the observation time. This allows us to speculate on the presence of an equilibrium associated with capacity of an organism to cope with health problems. That is, such equilibrium might indicate homeostatic limits when survivors begin to progressively loose their stress-resistance and accumulate avalanche-like health disorders. Actually, this is also reflected in sharp increase of mortality rates at moderate FI (see Figure 6). An interesting fact is that such limits for decedents correspond to 80-84 age group, i.e., the ages after which Gompertz model describing mortality age patterns begin to fail.

**Time to death (TTD), age, and frailty index (FI)**

To further assess the FI as a measure of BA and population heterogeneity, we next consider how individuals die. We begin from the analysis of the associations between FI, age, and TTD.

We calculated mean TTD for individuals of given age groups and for individuals with specified FI deciles (Figure 2, left panel). As it is expected, mean TTD for females is larger than for males. It is also seen that TTD declines with FI and age irrespective of sex. An intriguing finding is that magnitudes of the TTD for age groups and FI levels are similar, although they are slightly larger (especially for females) at younger ages than at older ages. It is evident that FI might be considered as an adequate measure of aging processes. Figure 2 (right panel) shows FI frequency distributions for each of the considered age groups. It is seen that gamma-distribution-like patterns persist up to the advanced ages (95 years). Consequently, proportion of individuals with small FI values remains relatively large by those ages, although it decreases with age. Therefore, from Figure 2 we can conclude that individuals with small FI might live almost as long as young elderly irrespective of their chronological age.

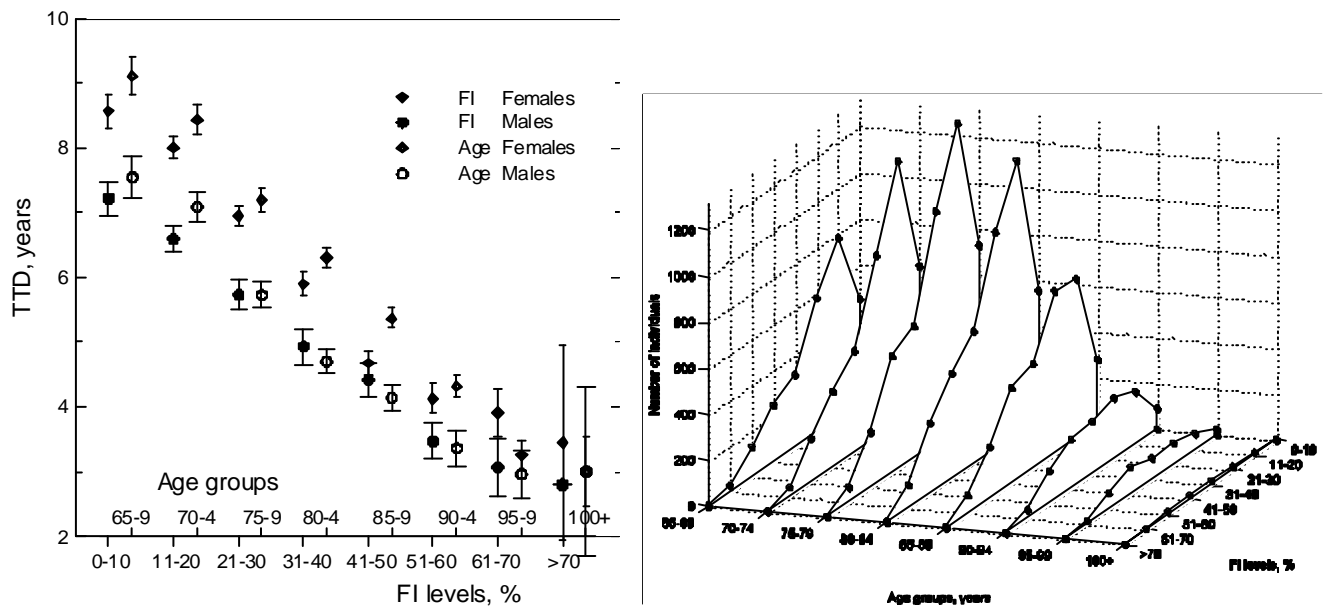

**Figure 2.** Left panel shows distribution of Time to death (TTD) by Frailty Index (FI) levels and age groups for both sexes of the PAD sample. Bars show 95% CI. Right panel depicts FI frequency distributions for each age group in the figure on left panel.

Despite the remarkable similarities between FI and age patterns seen in Figure 1 and Figure 2, chronological age and FI are largely independent measures of aging processes as we already inferred from their low correlation. Further support toward this view is provided by Figure 3 which shows FI and age distributions averaged over one year of TTD. Insights from this figure are consistent with our above conclusions, i.e., the FI can be considered as a biological indicator of aging and healthy individuals live longer. From Figure 3 nonlinear relations of FI and age with TTD for the entire PAD sample and for both sexes (except age-TTD pattern for males) are evident. However, the mean FI and age do not change with TTD in a similar fashion, i.e., the FI-TTD relationship exhibits concave pattern while the age-TTD pattern is convex. Therefore, the same increase in TTD leads to a sharper decline in mean FI than in age. Consequently, the FI better captures population heterogeneity with respect to death than chronological age does.

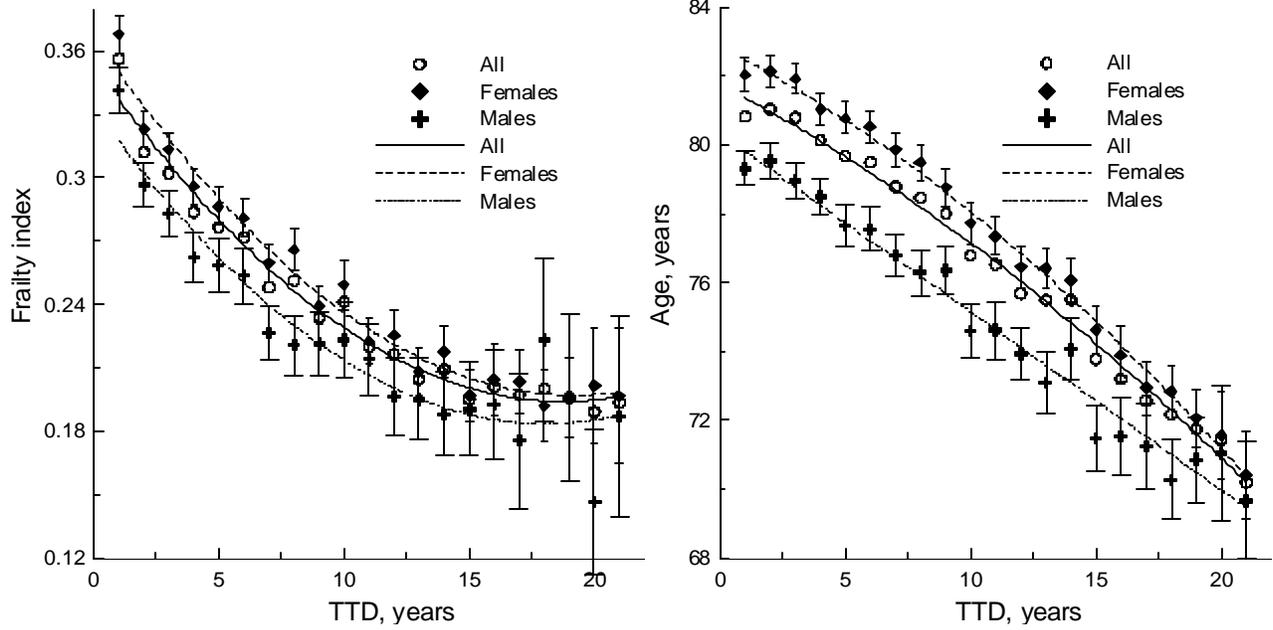

**Figure 3. Time to death (TTD) patterns of the Frailty Index (FI) (left panel) and chronological age (right panel) for whole PAD sample (all) and for both sexes. Bars show 95% CI for males and females.**

To find the best fit to the FI and age patterns we tested four functions: linear, exponential, power-law, and quadratic. Among them, the best fit was quadratic, $y = U + B_1 \times TTD + B_2 \times TTD^2$, where $y$ is FI or age. However, coefficient $B_2$ for the fit of age-TTD pattern for males was insignificant thus indicating that a better fit is given by linear function. The coefficients for each pattern are in Table 1. For male and female subgroups the difference between the coefficients $B_{1,2}$ for quadratic function is statistically insignificant. Therefore they should be considered equal. Also, the slopes of these fits, ($B' = B_1 + B_2 \times TTD$) and their variation with TTD ($B_2$), appear to be similar. Consequently, the TTD-patterns of the FI for males are nearly parallel but shifted down with respect to the female pattern since male and female intercepts ($U$) are statistically different.

**Table 1. Coefficients for quadratic and linear fits in Figure 3 and coefficients of determination.**

| Pattern | Sample | $B_1\ (SE) \times 10^2$ | $B_2\ (SE) \times 10^4$ | $U\ (SE)$ | $R^2$, % |
|---|---|---|---|---|---|
| FI-TTD | PAD | -1.70 (0.11) | 4.53 (0.50) | 0.354 (0.005) | 97.7 |
|  | Males | -1.66 (0.24) | 4.61 (1.07) | 0.334 (0.012) | 89.0 |
|  | Females | -1.78 (0.12) | 4.63 (0.55) | 0.367 (0.006) | 97.6 |
| Age-TTD | PAD | -0.382 (0.052) | -0.008 (0.002) | 81.77 (0.25) | 99.1 |
|  | Males[#] | -0.517 (0.020) | - | 80.33 (0.25) | 97.2 |
|  | Females | -0.379 (0.046) | -0.010 (0.002) | 82.87 (0.22) | 99.4 |

Note: Numbers in parentheses are standard errors (SE); p<.001; [#] linear fit.

While the foregoing analysis indicate that FI might be an adequate BA indicator and measure of population heterogeneity in mortality models, it is not clear yet whether FI is a better indicator of aging than chronological age. Evidence that this is the case would be given by showing that the FI is better correlated with TTD than age with TTD. Table 2 shows total and sex-specific coefficients of correlation between FI and TTD, and age and TTD. As can be seen, the correlation coefficients $r_{FI-TTD}$ are, indeed, larger than $r_{Age-TTD}$ coefficients for shorter observation periods for the entire sample and for male/female subgroups. Actually, this is consistent with view of dynamical nature of frailty (Mitnitski et al. 2004, Fried et al. 2004). That is, the disregulation of multiple physiological systems will likely results in faster health decline than with normal aging and would be more pronounced for short-time periods. This will result in increased mortality for such individuals, and, consequently, in larger correlation of the FI than age with TTD for shorter observation periods.

Table 2. Correlation coefficients between FI and TTD ($r_{FI-TTD}$) and between age and TTD ($r_{Age-TTD}$).

| Observation period, yrs. | $r_{FI-TTD}$ | | | $r_{Age-TTD}$ | | |
|---|---|---|---|---|---|---|
| | PAD | Males | Females | PAD | Males | Females |
| 1 | -0.168 | -0.197 | -0.147 | -0.026[#] | 0.021[#] | -0.063* |
| 4 | -0.168 | -0.187 | -0.164 | -0.038 | -0.044* | -0.053 |
| 8 | -0.205 | -0.229 | -0.201 | -0.108 | -0.135 | -0.115 |
| 21 | -0.285 | -0.270 | -0.306 | -0.305 | -0.306 | -0.332 |

[#]$p>.05$, *$p<.05$, $p<.01$

**FI Age Patterns For Deceased And Survived Individuals**

The above results show the potential of the FI for capturing hidden heterogeneity of individual aging and its correlation with TTD. Another manifestation of these FI properties is nonlinear nature of the FI age patterns (Mitnitski et al. 2004, Kulminski et al. 2005) for decedents and survivors.

Figure 4 shows sex-specific FI age patterns for individuals who survived (left panel) and died (right panel) during one, four, and eight years after interview averaged for five-years of age. An immediate observation is that the FI age patterns for both survivors and decedents are of nonlinear nature, i.e., health deficits are accumulated at different rates for distinct ages. Nonlinearity of the FI age patterns seems to be their inherent feature, which does not depend either on vital status or sex. Although a typical view on BA markers is that they should exhibit linear relation with chronological age (Karasik et al. 2005), the plasticity of mortality at advanced ages and age-dependence of mortality rate variation at those ages in experiments with anti-aging interventions (Vaupel, Carey, and Christensen 2003) provide solid arguments toward nonlinear view on BA indices.

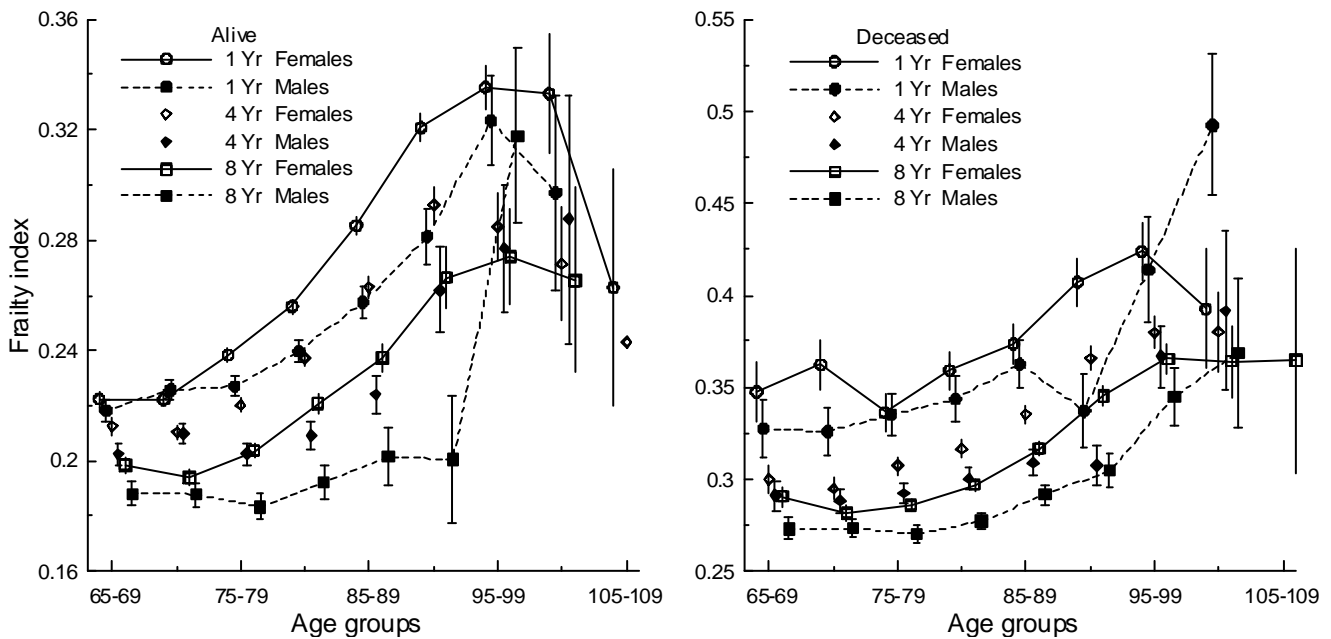

**Figure 4. Five-year age patterns of the Frailty Index for males and females survived (left panel) and died (right panel) during one, four, and eight years after interview. The standard errors ($\pm SE$) of means are shown by bars.**

**Markers without bars show imprecise results.**

FI age patterns show tendency to decelerate and even decline (especially for females) at advanced ages except for males dying (surviving) during one (eight) year(s). Moreover, upper bound of the SE for one year survivors also declines. We note that deceleration and decline are also characteristic features of mortality age patterns (see discussion after Figure 5). This fact provides further support to consider FI as a BA indicator and measure of hidden heterogeneity.

In general, females have larger FIs than males both for groups of survivors and decedents. This difference is statistically significant for surviving males and females aged 75-89 and for male and female decedents aged 90-94 irrespective of length of observation period. In addition, a statistically significant difference is seen for one-year survivors aged 90-94 as well as for those dying during four or eight years aged 85-89 and those dying during eight years aged 75-84. At younger ages the difference between mean FIs for males and females is small – especially for survivors. More interestingly, it is also diminishes irrespective of time under observation when males and females approach extreme ages (95+).

Figure 4 clearly shows that individuals with smaller number of deficits live longer irrespective of age, confirming our above conclusion. However, this conclusion does not cover all peculiarities of the age-pattern behaviors with time. First we note a nonlinear downward shift of the age pattern for deceased individuals with an increased observation period. That is, the difference between FIs for individuals died during one and four years is consistently larger than the difference between FIs for individuals died during four and eight years. Then we note nearly flat age pattern for eight-year male survivors and one-year male decedents. In fact, these features, along with those discussed above, might reflect physiological basis of aging process which is discussed in Section "Discussion and Conclusions."

**FI- And Age- Specific Mortality Rates (MR)**

To ascertain if observations made above are manifested in mortality patterns, we calculated age- and FI- specific mortality rates (MR). However, one-year MR estimates have a lack of statistical power at extreme ages (105+) and large FI level (>70%). Consequently, for better resolution of the observed tendencies at those ages and FI levels, we also calculated MR on the basis of deaths that occurred during four years of follow-up (four-year MR). Figure 5 shows age-specific one-year (left panel) and four-year (right panel) MRs for the entire PAD sample as well as for males and females.

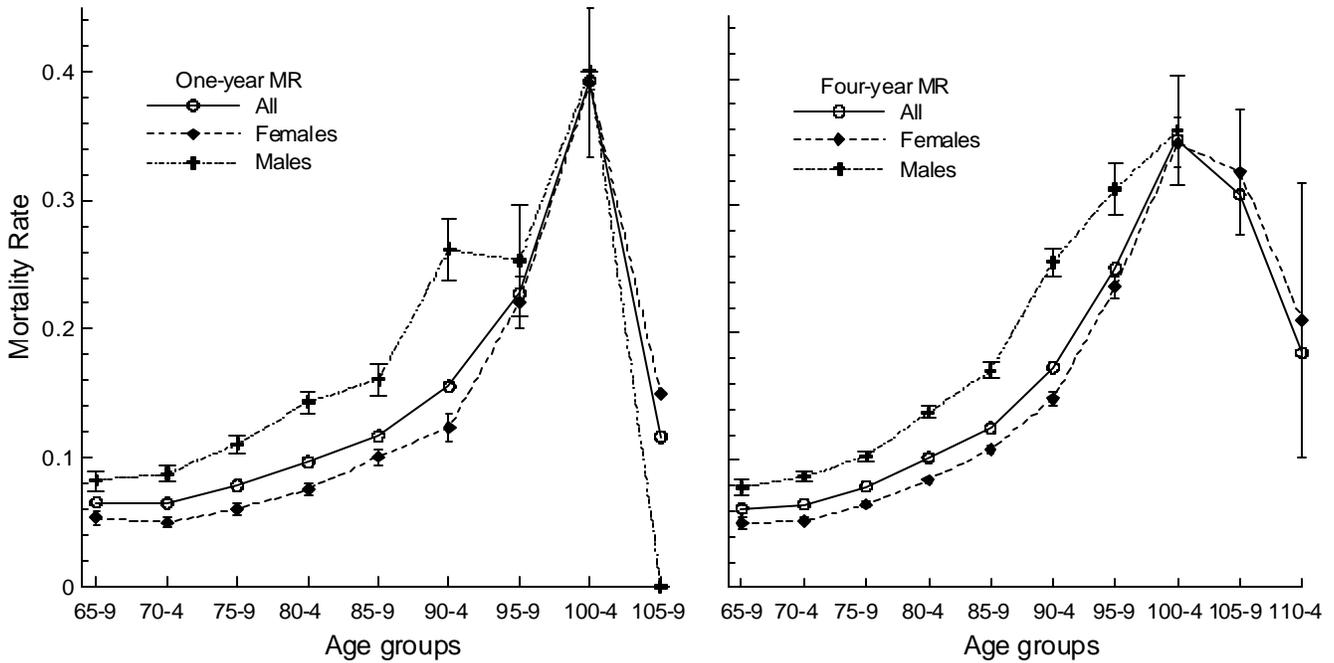

**Figure 5. Age patterns of the Mortality Rate (MR) for the entire PAD sample (all), as well as for males and females. Left (right) panel is plotted for one (four) year MR. Bars show $\pm SE$.**

It is seen that MRs exhibit nonlinear increase up to extreme ages for all samples. For females, this increase is generally accelerates up to age 100 years, as is clearly visible from the increasing angle of tangents to four-year MR patterns. For males, four-year MR patterns show consistent deceleration after age 90. For all samples, the one-year MR patterns show a tendency to decline at extreme ages (100+). This tendency becomes clearly pronounced for four-year MR patterns. Moreover, we observe that the upper bounds of the standard errors also decline with age, thus making our conclusion on the MR decline at extreme ages more reliable. In fact, all these tendencies are consistent with recent findings from MR patterns estimated for the U.S. elderly population (Manton et. al. 2005). Recall again that the patterns in Figure 5 pertain to MR among disabled elderly from the NLTCS sample. Therefore, it is not surprising that the MR for this sample are larger than for the general U.S. elderly population.

Analysis of Figure 4 and Figure 5 shows remarkable similarity in the behavior of the FI and MR with age. Specifically, FI age-patterns are not only similar to MR in terms of accelerated increases with age, but they also capture tendencies to decline at advanced ages. The direct association of the FI with MR can be evaluated from FI-specific MR. Since such patterns might be age-sensitive, we calculated sex-specific patterns averaged for all ages (Figure 6, left panel) as well as distinguishing younger individuals (65-84) and oldest-old (85+) (Figure 6, right panel).

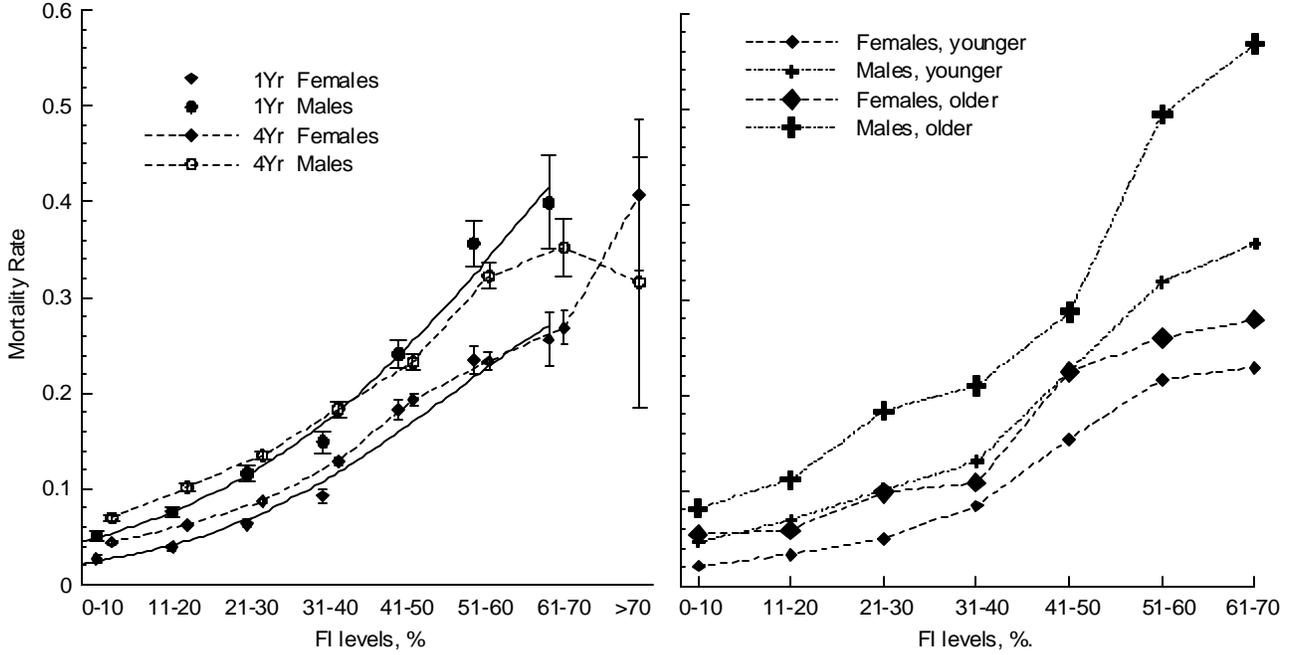

**Figure 6.** Left panel shows one (filled markers) and four (open markers) year FI-specific Mortality Rate (MR) patterns for males (dots) and females (rhombuses). Continuous line shows logistic fit for one-year patterns (see text for coefficients). Right panel shows one-year MR patterns for individuals of younger ages (65-84) (thin lines and small markers) and for the oldest-old (85+) (thick lines and large markers). Bars show $\pm SE$.

Figure 6 shows that, generally, MR tends to increase with FI. Again, we see that this increase is of nonlinear nature with small acceleration/deceleration at small/large FI and with faster (larger acceleration) increase at moderate FI. Fast increase of the MR at moderate FI is consistent with our conclusion from the analysis of the frequency patterns. Specifically, moderate FI levels might indicate the limits of organism's homeostatic capacity. Such nonlinear behavior is typical for a logistic growth model which describes population development (Strogatz 2000). For our case respective equation takes the form

$$MR = \left(K^{-1} + B_0 B_1^{FI}\right)^{-1}, \qquad (1)$$

where coefficients $K$, $B_0$, and $B_1$ need to be estimated from the data. Indeed, in tests of exponential, quadratic, power, and logistic functions to the data of Figure 6, we found that the latter one better fits the data for FI<70%. For the last FI level (>70%) the one- and four-year MR estimates are imprecise because of insufficient number of subjects (39 females with 12 death events and 16 males with four death events) for that level. Fits of FI-specific one-year MRs show that logistic function slightly better characterize the males' patterns than the females' ($R^2$=99%, $B_0$=30.5 [SE=2.86], $B_1$=0.0093 [SE=0.0019] at $K$=0.8 for males and $R^2$=98%, $B_0$=70.1 [SE 11.5], $B_1$=0.0029 [SE=0.0011] at $K$=0.4 for females). We can also see in this figure that the mean MR for males is typically larger than for females for a given FI level. This difference is statistically significant for age-averaged estimates except for FI>60% (left panel).

Figure 6 (right panel) also shows that younger individuals have smaller mean MR than older persons for the same level of FI. We also observe divergence of these patterns for males and females as FI increases, i.e., for larger FI this difference is larger than for smaller FI. Such behavior might be attributed to decline in stress resistance of an organism, which occurs with age (Semenchenko et al. 2004). The most noticeable difference is seen between younger and older males. This implies that males have smaller reserves to cope with stresses than females, i.e., smaller homeostatic capacity. The decline in stress resistance for males and females is enhanced by the nonlinear increase of the difference between respective MRs. That is, loss of stress-resistance is not only sex- and age- sensitive but also sensitive to magnitude of FI – the larger the FI is, the quicker disabled individuals can loose their stress-resistance.

**Relative Risks of Death**

Additional quantitative insights on connections of age and FI with mortality can be obtained by analyzing the relative risk (RR) of death. To analyze RR, we performed multivariate Cox regression analysis with age and FI as basic covariates and with and without control for sex, as well as for males and females separately. Table 3 shows the results of our analysis considering occurrence of death within one, four, eight years and whole observation period (21 years). We found that, for the short time periods under observation, the RR of death because of FI ($RR_{FI}$) is considerably larger than the RR of death because of age ($RR_{Age}$). Indeed, for one-year age increment $\ln(RR_{Age})$ is 0.032 while for 1% of FI increase $\ln(RR_{FI})$ is 0.039. Therefore, the one-year risk increment is equivalent to 0.82% (=0.032/0.039) risk increment due to FI resulting in the same RR of death. Consequently, the integrated risk (IR) of death due to accumulated deficits (from 0 to 100%) is 13.8 times larger than because of age for 40 years (age 65 to 105 years), i.e., $e^{0.039^{100}} / e^{0.032^{40}} \approx 13.8$. With increasing the observation periods, this ratio significantly decreases going to nearly unity (9.72/9.18=1.06) for the entire period. Nevertheless, the IR associated with FI remains larger than the IR associated with age.

**Table 3. Cox proportional multivariate hazard regression.**

| Observation period, yrs. | Covariate | ln(RR) | | | |
|---|---|---|---|---|---|
| | | PAD | | Males | Females |
| | | SEX | No SEX | | |
| 1 | AGE | 0.032 (0.003) | 0.027 (0.003) | 0.034 (0.004) | 0.031 (0.004) |
| | FI, % | 0.039 (0.001) | 0.039 (0.001) | 0.037 (0.002) | 0.042 (0.002) |
| 4 | AGE | 0.046 (0.001) | 0.041 (0.001) | 0.047 (0.002) | 0.045 (0.002) |
| | FI, % | 0.029 (0.001) | 0.029 (0.001) | 0.028 (0.001) | 0.031 (0.001) |
| 8 | AGE | 0.052 (0.001) | 0.047 (0.001) | 0.051 (0.002) | 0.052 (0.001) |
| | FI, % | 0.026 (0.001) | 0.026 (0.001) | 0.025 (0.001) | 0.027 (0.001) |
| 21 | AGE | 0.056 (0.001) | 0.051 (0.001) | 0.054 (0.002) | 0.056 (0.001) |
| | FI, % | 0.023 (0.000) | 0.022 (0.000) | 0.023 (0.001) | 0.023 (0.001) |

Note: Numbers in parentheses are standard errors.
p<.001

Without controlling for sex (Table 3, "No SEX" column) $\ln(RR_{FI})$ remains practically the same. Meanwhile, $\ln(RR_{Age})$ slightly decreases when sex is controlled. Surprisingly, the $\ln(RR_{FI})$ for males is slightly smaller than for females indicating that RR for females per 1% of the FI increase is larger than for males (i.e., that RR for females is larger than for males). This observation is consistent with findings of a recent study on sex differences in the risk of death among frail elderly in the Longitudinal Aging Study of Amsterdam (Puts et al., 2005). This is likely due to the facts that females both accumulate more deficits than males over their life and live longer. Interplay among these two tendencies might result in the fact that their RR of death can be larger per unit of FI while IR/RR for the same FI can be smaller than for males. Indeed, this conclusion can be confirmed by considering RR for the same FI levels for males and females. To do this we selected four FI levels: FI(0) ($0 \le FI \le 10\%$), FI(1) ($11 \le FI \le 20\%$), FI(2) ($21 \le FI \le 30\%$), and FI(3) ($FI > 30\%$).

Table 4 shows the results for both sexes for short (1 year) and long (21 years) observation periods. The RR is calculated with respect to the baseline hazard at FI(0). It is seen that for whole observation period the RR for males for the same FI level is consistently larger than for females. The largest difference occurs for FI(2) and the smallest for FI(3), although they are not statistically significant. For the short observation period, the RR for males is larger than that for females for small FI; becomes comparable for moderate FI; and becomes larger for large FI. For all FI levels, the RR for both sexes diminishes as observation time increases. The largest statistically significant difference between RRs for one and 21 years of observation is seen for females for FI(3). This fact might be associated with essential contribution of "survivors" as discussed in the following section.

**Table 4. Cox proportional multivariate hazard regression for males and females observed during short (1 year) and long (21 years) time periods.**

| Observation period, yrs. | SEX | ln(RR) | | | |
|---|---|---|---|---|---|
| | | Age | FI(1) | FI(2) | FI(3) |
| 1 | Males | 0.033 (0.004) | 0.383 (0.129) | 0.781 (0.130) | 1.442 (0.114) |
| | Females | 0.033 (0.004) | 0.322 (0.144)* | 0.769 (0.140) | 1.609 (0.130) |
| 21 | Males | 0.054 (0.002) | 0.278 (0.036) | 0.512 (0.039) | 0.855 (0.036) |
| | Females | 0.057 (0.001) | 0.194 (0.031) | 0.416 (0.032) | 0.805 (0.031) |

Note: Numbers in parentheses are standard errors.
*p<.05, p<.001

Since age and FI exhibit nonlinear relation with TTD and MR, their impact on the RR might be age and FI dependent (nonlinear) as well. To assess the possible nonlinearity, we performed the similar analysis as for Table 4 but for five-year age groups and for FI deciles. The first age group (65-69) and the 0-10% FI level were considered as a baseline. Figure 7 summarizes this analysis showing ln($RR_{Age}$) (left panels) and ln($RR_{FI}$) (right panels) for the PAD sample, controlling for sex as well as for males and females separately for four (upper panels) and 21 (lower panels) years of follow up.

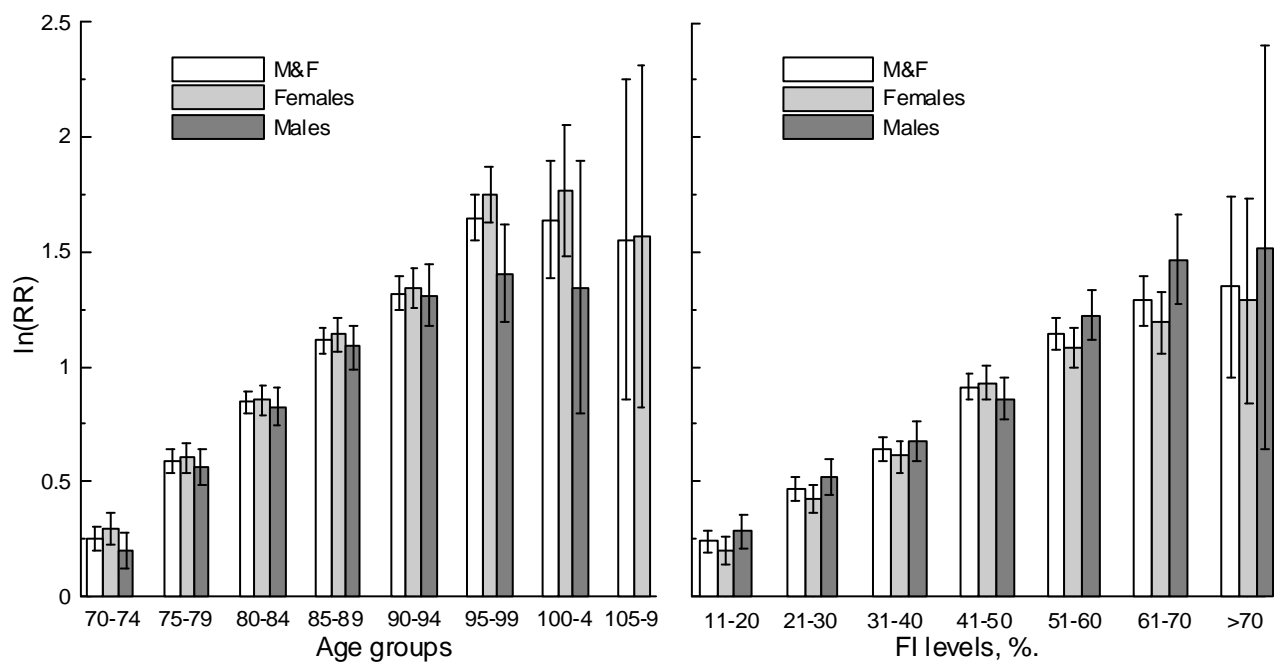

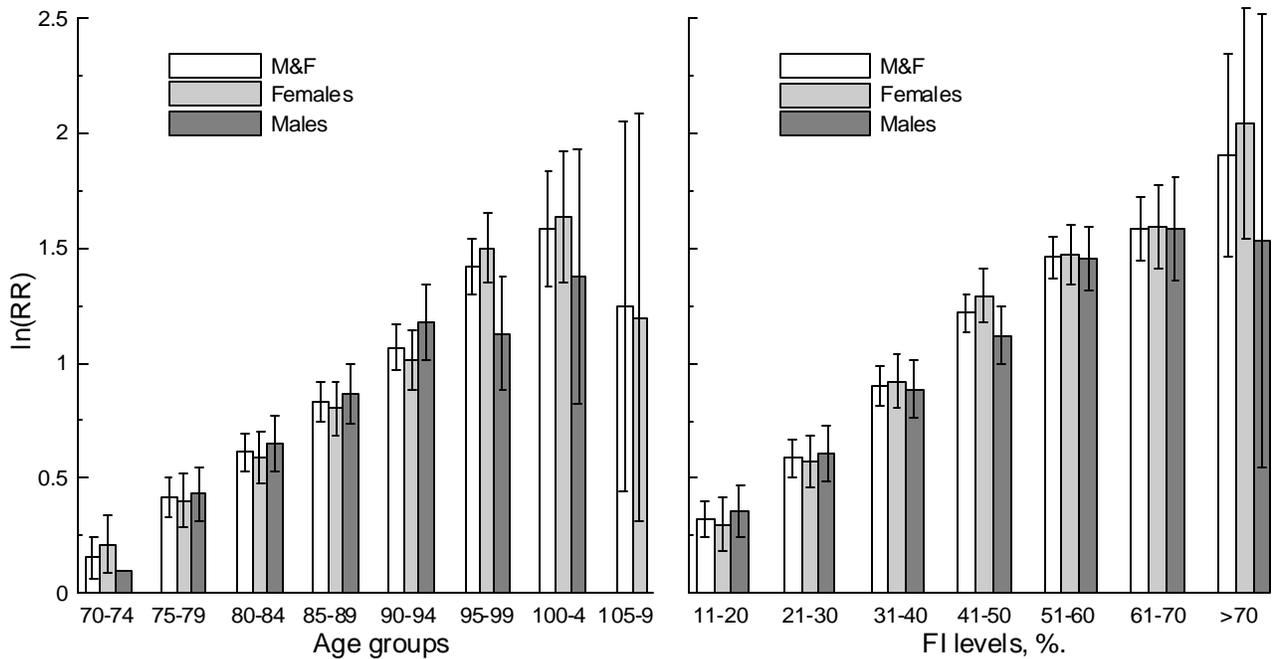

**Figure 7. Logarithm of Relative Risk (RR) of death for different five-year age groups (left panels) and for the Frailty Index (FI) deciles (right panels) for the entire PAD sample (white bars), females (light-grey), and males (dark-grey) for four- (upper panels) and 21- (lower panels) year observation periods. Thin bars show 95% CI. For thick bar without thin bar the estimate is insignificant.**

It is seen that $\ln(RR_{Age})$ for the PAD sample increases linearly with age until advanced ages (95-99 age group) and exhibits a decline after that age. The decline with age is also seen for both sexes. Deviation from the linear growth of the exponential factor, i.e., $\ln(RR)$, at the oldest-old ages means that RR is not an exponential function of age and exhibits decline in RR by extreme ages (100+). For the entire observation period $\ln(RR_{FI})$ patterns show similar behavior for all samples. For shorter observation time $\ln(RR_{FI})$ patterns for females and the entire sample exhibit nearly linear growth ($R^2=98.3$ and $R^2=99.0$ for females and the entire sample, respectively, vs. $R^2=94.3$ for males). Our analysis does not show statistically significant differences in the RR of death for males and females, except for 95-99 age group. Thus, again, we see similarity between RR age- and FI- patterns.

**DISCUSSION AND CONCLUSIONS**

We performed a detailed analysis of statistical properties of a cumulative index of age-associated disorders (deficits), called a "frailty index" (FI), in order to assess the possibility that the FI can describe aging processes, population heterogeneity, and predict death in the elderly. We focused our analysis on disabled individuals who are at excessive risk of death and, consequently, for whom all these aspects of the FI are more pronounced. We comprehensively addressed the problem of sex-sensitivity of the FI in disabled elderly. We also showed that the FI can capture a number of physiological processes underlying aging phenomenon.

Our analysis reveals remarkable similarities between various statistical characteristics of FI and age. Specifically, we found that the frequency, time-to-death, mortality-rate, and relative-risk-of-death distributions over the FI and chronological age are remarkably similar. This similarity is seen not only for the entire sample but also for males and females. Moreover, it is also preserved for surviving and deceased individuals. One reason for such similarities could be a high correlation of the FI and age. We showed that this is clearly not the case – while this correlation is statistically significant, it is relatively small being not able to explain such similarities. All these findings provide strong evidence that the FI is an adequate measure of the aging process and that it is distinct from chronological age.

An intriguing finding is that the FI, on the one hand, and mortality rates and relative risk of death, on the other hand, exhibit closely similar age patterns with accelerated increases up to oldest-old ages and with subsequent deceleration and even decline. We note that deceleration and decline in mortality rates at extreme ages is a characteristic feature of short-lived experimental species (Vaupel et al. 1998; Carey et al. 1992). These features were also noted in human population (Vaupel et al. 1998). Recently, a comprehensive analysis of mortality rates in the U.S. elderly provided additional evidence on their deceleration/decline at extreme ages (Manton et al. 2005). Therefore, the fact that the FI age patterns exhibit tendencies of accelerated increase at younger ages, deceleration at advanced ages, and even decline at extreme ages (especially for females), which are similar to those seen for the mortality-rate and relative-risk-of-death age patterns, provides solid evidence for the FI as a reliable indicator of aging in population-based studies of mortality and aging. Moreover, our analysis shows that the FI is a better indicator of aging than chronological age. Since TTD-FI patterns exhibit concave shape while age-TTD pattern is convex, we also argue that FI better captures population heterogeneity than chronological age.

Analysis of various FI patterns shows that the FI can likely capture physiological processes underlying aging both on individual and population levels. In particular, we show that the FI can be associated with the limits of an organism's homeostatic capacity and be a characteristic of the organism's stress-resistance. The nonlinear downward shift of the FI age pattern with observation time for decedents might be a manifestation of the compression of morbidity (Fries 1983, 2000, 2003), since it means that accumulation of deficits occurs with faster rate prior to death. Meanwhile, decelerating and even declining rates of deficit accumulation manifested above the oldest-old ages as well as a convergence of the age patterns for deceased females distinguishes this situation from the compression of morbidity scenario and makes it more optimistic. That is, individuals can live relatively long at extreme ages (e.g., 4 years, 8 years) with a large FI. This conclusion is also supported by the analysis of the TTD patterns.

The fact that certain individuals at extreme ages in poor health die fast while the others live long was noted by Evert et al. (2003) and has lead to the identification of three groups of centenarians with markedly different mortality profiles (Evert et al. 2003; Andersen et al. 2005). The first group is "survivors", i.e., those individuals who were unhealthy before age 80. The second group is "delayers" (those being unhealthy from 80 to 100), and the last group is "escapers" (no major age-related lethal diseases before age 100). Our analysis (see Figure 4) shows that individuals at extreme ages who remain alive 4 and 8 years after the interview date have considerably smaller mean FI than individuals who die in one year. This fact allows us to speculate about heavy prevalence of "delayers" or "escapers" among centenarians in the NLTCS. In addition, the relatively large mean FI for individuals dying during a long period of time indicates the presence of a significant portion of "survivors".

Our analysis clearly shows that males surviving during the same time period as females should be healthier. It is worth noting that, as the time under observation increases, the rate of FI increase with age for males declines, and, finally, for long time periods, FI age patterns become flat at 65-94 years irrespective of vital status. However, the FI for males sharply increases at 95+ becoming as large as that for males surviving a short time and nearly the same for those died during eight years. This observation might be associated with biological age limits, which characterize the level of health-maintenance in the society (Kulminski et al. 2005, Fogel 1997) and does not mean that longevity cannot be extended beyond certain age. Such behavior is also compatible with the compression of morbidity. Then, presence of the BA limits provides a biological explanation thereof.

In conclusion, the FI is an adequate indicator of aging and population heterogeneity suitable for various models of aging and mortality. Actually, FI appears to be a better indicator than chronological age. The FI also appears to be a measure which captures physiological processes underlying aging on an individual and population levels.

# REFERENCES

Andersen, S.L., D.F. Terry, M.A. Wilcox, T. Babineau, K. Malek, and T.T. Perls. 2005. "Cancer in the Oldest Old." *Mechanisms of Ageing and Development* 126(2):263-7.

Beard, R. 1971. *Some Aspects of Theories of Mortality Cause of Death Analysis, Forecasting and Stochastic Processes*. In: Brass, W. (Ed.). Biological Aspects of Demography. London: Taylor & Francis, 557–68.

Bennett, A.K. 2004. "Older Age Underwriting: Frisky vs Frail." *Journal of insurance medicine* 36(1):74-83.

Bortz, W. 2nd. 2002. "A Conceptual Framework of Frailty: a Review." *The Journal of Gerontology. Series A, Biological Sciences and Medical Sciences* 57:M283-8.

Buchner, D.M., E.H. Wagner. 1992. "Preventing Frail Health." *Clinics in geriatric medicine* 8:1-17.

Carey J.R., P. Liedo, D. Orozco, and J.W. Vaupel. 1992. "Slowing of Mortality Rates at Older Ages in Large Medfly Cohorts." *Science* 258:457–61.

Council on Scientific Affairs. 1990. American Medical Association White Paper on Elderly Health. Report of the Council on Scientific Affairs. *Archives of Internal Medicine* 150:2459-72.

Evert J., E. Lawler, H. Bogan, and T. Perls. 2003. "Morbidity Profiles of Centenarians: Survivors, Delayers, and Escapers." *The Journal of Gerontology. Series A, Biological Sciences and Medical Sciences* 58(3):232-7.

Fogel, R. 1997. "Economic and Social Structure for an Ageing Population. *Philosophical transactions of the Royal Society of London. Series B, Biological sciences* 352:1905-17.

Fretwell, M. 1993. *Acute hospital care for frail older patients*. In: Hazzard, W.R., E.L. Bierman, J.P. Blass, W.H.J. Ettinger, and J.B. Halter, eds. *Principles of Geriatric Medicine and Gerontology*. 3rd Ed. New York: McGraw-Hill; 241–248.

Fried, L.P., C.M. Tangen, J. Walston, A.B. Newman, C. Hirsch, J. Gottdiener, T. Seeman, R. Tracy, W.J. Kop, G. Burke, and M.A. McBurnie. 2001. "Frailty in Older Adults: Evidence for a Phenotype." *The Journal of Gerontology. Series A, Biological Sciences and Medical Sciences* 56A:M146-M156.

Fried L.P., and J. Walston. 2003. "*Frailty and Failure to Thrive.*" In: *Principles of Geriatric Medicine and Gerontology*. 5th Ed. Hazzard, W.R., J.P. Blass, W.H. Ettinger, J.B. Halter, J. Ouslander, eds. New York: McGraw-Hill; 1487–502.

Fried, L., L. Ferrucci, J. Darer, J. Williamson, and G. Anderson. 2004. "Untangling the Concepts of Disability, Frailty, and Comorbidity: Implications for Improved Targeting and Care." *The Journal of Gerontology. Series A, Biological Sciences and Medical Sciences* 59:255-63.

Fries, J.F. 1983. "The Compression of Morbidity." *Annals of the Academy of Medicine, Singapore* 12(3):358-67.

------. 2000. "Compression of Morbidity in the Elderly." *Vaccine* 18(16):1584-9.

------. 2003. "Measuring and Monitoring Success in Compressing Morbidity." *Annals of Internal Medicine* 139(5 Pt 2):455-9.

Gillick, M. 2001. "Pinning Down Frailly." *Journal of Gerontology. Series A, Biological Sciences and Medical Sciences* 2001; 56A: M134-M135.

Hogan, D.B., C. MacKnight, and H. Bergman. 2003. "Models, Definitions, and Criteria of Frailty." *Aging clinical and experimental research* 15(3 Suppl):1-29.

Jazwinski, S.M. 2002. "Biological Aging Research Today: Potential, Peeves, and Problems." *Experimetal Gerontology* 37(10-11):1141-6.

Karasik, D., S. Demissie, L. Cupples, and D. Kiel. 2005. "Disentangling the Genetic Determinants of Human Aging: Biological Age as an Alternative to the Use of Survival Measures." *The Journal of Gerontology. Series A, Biological Sciences and Medical Sciences* 60:574-87.

Kirkwood, T.B.L. 2002. "New Science for an Old Problem." *Trends in genetics* 18:441–2.

Kowald, A. 2002. "Lifespan Does Not Measure Ageing." *Biogerontology* 3(3):187-90.

Kulminski, A., A.I. Yashin, S.V. Ukraintseva, I. Akushevich, K.G. Arbeev, K.C. Land, and K.G. Manton. 2005. "Age-associated disorders as a proxy measure of biological age: Findings from the NLTCS Data." Preprint # q-bio/0509034 published at http://arxiv.org/abs/q-bio/0509034.

Lipsitz, L.A. and A.L. Goldberger. 1992. "Loss of "Complexity" and Aging. Potential Applications of Fractals and Chaos Theory to Senescence." *The journal of the American Medical Association* 267:1806-1809.

Manton, K.G., I. Akushevich, and A. Kulminski. 2005. "Human Mortality at Extreme Ages: New Data and Analysis." Submitted to Demography.

Manton, K.G., and A. Yashin. 2000. *Mechanisms of Aging and Mortality: Searches for New Paradigms.* Odense, Denmark: Odense University Press.

Markle-Reid, M. and G. Browne. 2003. "Conceptualizations of Frailty in Relation to Older Adults." *Journal of advanced nursing* 44(1):58-68.

Mitnitski, A.B., J.E., Graham, A.J. Mogilner, and K. Rockwood. 2002. "Frailty, Fitness and Late-Life Mortality in Relation to Chronological and Biological Age." *BMC geriatrics* 27:2(1):1-8.

Mitnitski, A.B., X. Song, and K. Rockwood. 2004. "The Estimation of Relative Fitness and Frailty in Community-Dwelling Older Adults Using Self-Report Data." *The Journal of Gerontology. Series A, Biological Sciences and Medical Sciences* 59(6):M627-32.

Mitnitski, A., A. Mogilner, and K. Rockwood. 2001. "Accumulation of Deficits as a Proxy Measure of Aging." *Scientific World Journal* 1:323-36.

Morley, J.E., H.M. III Perry, and D.K. Miller. 2002. "Something About Frailty." *The Journal of Gerontology. Series A, Biological Sciences and Medical Sciences* 57A:M698-M704.

Puts, M.T., P. Lips, and D.J. Deeg. 2005. "Sex Differences in the Risk of Frailty for Mortality Independent of Disability and Chronic Diseases." *Journal of the American Geriatric Society* 53(1):40-7.

Rockwood, K., D.B. Hogan, and C. MacKnight. 2000. "Conceptualization and Measurement of Frailty in Elderly People." *Drugs Aging* 17:295-302.

Rockwood, K., A. Mogilner, and A. Mitnitski. 2004. "Changes With Age in the Distribution of a Frailty Index." *Mechanisms of Ageing and Development* 125(7):517-9.

Rockwood, K., X. Song, C. MacKnight, H. Bergman, D.B. Hogan, I. McDowell, and A. Mitnitski. 2005. "A Global Clinical Measure of Fitness and Frailty in Elderly People." *Canadian Medical Association journal* 173(5):489-95.

Rockwood, K., A.B. Mitnitski, and C. MacKnight. 2002. "Some Mathematical Models of Frailty and Their Clinical Implications." *Review of Clinical Gerontology* 12:109–17.

Semenchenko, G., A. Khazaeli, J.W. Curtsinger, and A Yashin. 2004. "Stress Resistance Declines With Age: Analysis of Data From Stress Experiments With Drosophila Melanogaster. *Biogerontology*. 5(1):17-30.

Strogatz, S. 2000. *Nonlinear Dynamics and Chaos: With Applications to Physics, Biology, Chemistry and Engineering*. Perseus Books Group.

Vaupel, J., J. Carey, and K. Christensen. 2003. "It's Never Too Late." *Science* 301:1679-81.

Vaupel, J., J. Carey, K. Christensen, T. Johnson, A. Yashin, N. Holm, I. Iachine, V. Kannisto, A. Khazaeli, P. Liedo, V. Longo, Y. Zeng, K.G. Manton, and J. Curtsinger. 1998. "Biodemographic Trajectories of Longevity." *Science* 280:855-60.

Vaupel, J., K.G. Manton, and E. Stallard. 1979. "The Impact of Heterogeneity in Individual Frailty on the Dynamics of Mortality." *Demography* 16:439-54.

Woodbury, M. and K.G. Manton. 1977. "A Random Walk Model of Human Mortality and Aging." *Theoretical Population Biology* 11:37-48.

Yashin, A., K.G. Manton, and J. Vaupel. 1985. Mortality and Aging in Heterogeneous Populations: A Stochastic Process Model with Observed and Unobserved Variables. *Theoretical Population*

*Biology* 27:154-75.

Yates, F.E. 2002. "Complexity of a Human Being: Changes With Age." *Neurobiology of aging* 23(1):17-9.